\documentstyle[twoside,fleqn,espcrc2]{article}

\newcommand{\AmS}{{\protect\the\textfont2
  A\kern-.1667em\lower.5ex\hbox{M}\kern-.125emS}}

\def\Journal#1#2#3#4{{#1} {\bf #2}, #3 (#4)}

\def\NCA{\em Nuovo Cimento}

\def\NPB{{\em Nucl. Phys.} B}
\def\PLB{{\em Phys. Lett.}  B}
\def\PRL{\em Phys. Rev. Lett.}
\def\PRD{{\em Phys. Rev.} D}
\def\PRP{{\em Phys. Rept.}}

\def\be{\begin{equation}}
\def\ee{\end{equation}}
\def\bea{\begin{eqnarray}}
\def\eea{\end{eqnarray}}
\def\sla{\raise.15ex\hbox{$/$}\kern-.57em}

\title{
\vspace{-5.0cm}
\begin{flushright}{\normalsize RUHN-99-1}\\
\end{flushright}
\vspace*{2.5cm}
Chiral fermions on the lattice
\thanks{Talk at Lattice '99, June 29 - July 3, Pisa, Italy.}
}

\author{Herbert Neuberger\address{Department of Physics and Astronomy, \\ 
        Rutgers University, Piscataway, NJ 08855-0849}%
        \thanks{Research supported in part by the DOE, 
grant \#DE-FG05-96ER40559.}
        }
\begin{document}

\begin{abstract} 
Chiral fermions resisted being put on the lattice for twenty years.
This raised the suspicion that asymptotically free chiral gauge theories
were not renormalizable outside perturbation theory and therefore
could not be mathematically extended to infinite energies. During the
last several years the situation has reversed itself. Today we believe
that all the essential ingredients for a full lattice definition
of non-anomalous chiral gauge theories are in place within the overlap
construction. This construction is based on earlier
work by Callan and Harvey, by Kaplan and by Frolov and Slavnov. It can
be reinterpreted as coming from the Ginsparg-Wilson relation, but,
at the moment, it is a unique construction and therefore might
also be the way by which Nature itself regularizes chiral fermions.
This is yet another instance in which lattice field theory makes
a potentially important contribution to fundamental particle physics. 
\end{abstract}
\maketitle

\section{Introduction}
My talk is divided into three parts. In the first part the relevance
of the chiral fermion issue to fundamental particle physics and to
numerical QCD will be explained. The second part  is the
bulk part of my talk and will present the main ideas and properties
of the ``overlap''. I shall restrict myself to the period
from 1992 to one year ago, summer 1998, the time of the lattice 98
conference, held in Boulder, Colorado. Presumably the next plenary
speaker on chiral fermions will focus on the last year, between summer
1998 and summer 1999, so overlap will be avoided. In the last part
of my talk I shall try to convince you that these new developments
present many opportunities for fresh ideas.

My main partner in the overlap construction was R. Narayanan. I have
also collaborated with P. Huet, Y. Kikukawa, A. Yamada and P. Vranas.
Important contributions to the overlap development were made by
Randjbar-Daemi and J. Strathdee.
More recently, the pace of developments has picked up, mainly in the vector-like
context, and beautiful
work has been done by R. Edwards, U. Heller, J. Kiskis, by Ting-Wai Chiu
and by Keh-Fei Liu and his collaborators. 
New results are coming out almost daily, but this
is material for the next lattice conference. 

\section{Relevance}

The minimal standard model (MSM) works well experimentally, is a chiral
gauge theory and constitutes a good effective low energy description of the
theory of everything (TOE). The relatively varied $U(1)_Y$ 
charges of the MSM reflect
the chiral nature of the MSM by assuring anomaly free couplings
to the gauge bosons and gravity. These charges fit snuggly into 
representations of larger groups leading to $SU(5)$ and $SO(10)$ grand
unified theories (GUT). All these new gauge theories
are also chiral. Supersymmetric extensions of the MSM or of the
GUTs also must be chiral. By normal physics standards, the TOE is best
described as unknown at present. It seems likely that it is not
an ordinary field theory as it contains gravity. It could be the
case that in the TOE there is no well defined concept of chirality.

The most basic question about chirality was asked by Holger-Bech Nielsen
many years ago\cite{hbn_rome}. I paraphrase it to: Is a chiral gauge theory completely
isolated from ultra-violet effects ? In other words is it truly
renormalizable ? In yet another equivalent form the question is:
Is the chiral nature of the theory compatible with an 
arbitrarily large scale
separation between typical scales and new physics scales ?

We know very well that the answer to the above question is``yes" in
perturbation theory to any order. But, outside perturbation
theory, lattice difficulties have raised
the suspicion that the answer might be ``no". This was the situation
from the early 1970s to the early 1990s. The main achievement I am
reporting on occurred during the period 1992 to 1997 and amounts
to replacing the suspected ``no" by an almost compelling ``yes". 

Accepting the ``yes" from now on, the next question one should
ask is: What can we learn about Nature from the lattice difficulties
and from their resolution ? 

A mathematician might answer that we learn nothing because Nature isn't 
a lattice. This physicist's answer is different: If in Nature
there were an infinite number of fermions per unit four volume,
chirality at low energies can emerge naturally, without fine tuning.
The TOE could ``know" nothing about chiral gauge theories. New
mechanisms in the TOE produce the appropriate set of light
degrees of freedom in a natural way. Chiral gauge theories appear
because they are the single consistent interaction between these light
degrees of freedom in the long distance limit.

So, lattice field theory may have made a contribution to the
understanding of one of the most fundamental issues in theoretical
particle physics. This would certainly not be the first time,
but is worth keeping in mind in the social climate of today's 
particle theory. 

An important spinoff of the developments on chiral gauge theories
is feeding back into our own subfield. For the first time we
know, in principle and also in practice (to some degree), that
numerical QCD can treat global chiral symmetries exactly, on the
lattice. This subfield is rapidly expanding and I am sure 
more will happen over the next few years. 

\section{Overlap essentials}

\subsection{Infinite number of fermions}

If the number of fermions is infinite the theory is not 
precisely defined (yet). This provides an opportunity to
``cheat'':

We can write down a theory that looks vectorial, but could equally
well be viewed as chiral. Suppose we have a string of right-handed
Weyl fermions stretching from $-\infty$ to $+\infty$ and below it
another string, now made out of left-handed fermions. We can think
about the different fermions being labeled by a new discrete flavor
index increasing along the strings. If we pair right-handed
with left-handed Weyl particles starting from the middle of the strings
outwards, we conclude that we have an infinite number of Dirac fermions
and the theory is vector-like. However, suppose we start pairing from
both infinite ends inwards. One could easily make a ``mistake'' and
end up with, say, one  unpaired Weyl fermion. Now the theory looks
chiral.

We are restricting the action to be bilinear in the fermion fields.
The action for a multiplet of Weyl fermions would be $\bar\psi W \psi$,
where $W$ is the Weyl operator in a background gauge field. We shall
always assume we are working on a compact Euclidean manifold (it makes
no sense to have to worry also about infrared issues, the thermodynamic
limit and so forth - we have our plate full already). We know that
the gauge fields over a compact manifold fall into distinct ``blobs'',
each labeled by the topological charge $Q\{U\}$, 
where $U$ represents the gauge background in lattice notation 
although at this point we are still in the continuum. 
Moreover, we know that $W$ is structurally affected by 
$Q\{U\}$ via the famous Atiyah-Singer index theorem, which,
in a loose but quite sensible sense means that 
(the number of rows of $W$)-(the number of columns
of $W$)=$Q\{U\}$. Clearly, one cannot just write down
a finite size matrix $W$ with such a property. But, if the size
of $W$ is infinite it is at least conceivable that the Atiyah-Singer
index theorem will hold on the lattice since, after all, $\infty - \infty$
can be anything. 

It is easy to imagine writing down a formula for $W$ which is gauge 
covariant. This means that replacing the background gauge field
by a gauge transform is equivalent to conjugation by a formally unitary
matrix dependent only on the gauge transformation. Thus, $\det W$
would be gauge invariant. But, since $W$ is infinite, $\det W$ isn't
well defined and it is conceivable that
gauge invariance does not hold since the manipulations 
involving the determinant of an infinite matrix, 
the product of infinite
matrices and the only formal unitarity of some of these
may end up being incorrect. 
This creates an opportunity for anomalies to enter, an opportunity
we are obligated to create one way or another. It 
is important to see that anomalies can show up in a way that is
independent of taking the ultraviolet cutoff to infinity or the
infrared cutoff to zero. 

Thus, postulating an infinite number of fermions creates the right openings,
and the problem becomes only how to ``cheat'' honestly.

\subsection{Brief history of chiral issues}

In the late sixties Adler, Bell, Jackiw and Bardeen discovered
and understood anomalies in the context of particle physics \cite{basic}. 
The importance of this discovery cannot be overstated. 
Starting from the mid seventies Stora, Zumino \cite{stora_zumino} 
and others 
unveiled the beautiful mathematical structure of anomalies.
At the algebraic level the very elegant descent equations 
were seen to relate anomalies in various dimensions. During
the early to mid eighties the understanding of anomalies was
enriched by discovering their topological meaning. It is
fair to say that during this period the physics of fermions
in a classical gauge background was put on firm (and elegant)
mathematical grounds \cite{reviews}. 

For what follows, a crucial step was 
taken by Callan and Harvey \cite{callan_harvey}
who provided a physical realization of the relation 
between anomalies in different
dimensions (as algebraically reflected in the descent equations).
They connected the consecutive dimensions in the descent
equations by studying physical embeddings in a given manifold 
of sub-manifolds (``defects'') of lower
dimension. Prompted by this work, Boyanowski, Dagotto and Fradkin
\cite{boya} 
studied similar arrangements in condensed matter. They also
proposed that the famous chiral fermion problem of lattice field
theory might be solved this way. But, they did not pursue their
insight and our community paid no attention. 

The situation changed in 1992 when Kaplan\cite{kaplan}, again motivated by
the work of Callan and Harvey, made a very specific and compelling
case that the setup Callan and Harvey used could be put on the lattice
and that this was a new way to deal with the lattice fermion
problem. During the same year, starting from a completely different
point of view, Frolov and Slavnov \cite{frolov_slavnov} 
made a proposal containing an infinite number of auxiliary
fields to regulate $SO(10)$ gauge theory 
with a 16-plet of left-handed Weyl fermions. This is
a chiral gauge theory and each irreducible
matter multiplet can accommodate one family of the MSM plus one additional
$SU(3)\times SU(2)\times U(1)$-neutral right-handed neutrino. 

Narayanan and I 
asked whether the two ways, one by Kaplan, and the
other by Frolov and Slavnov had anything in common. Our conclusion
was that they did and the heart of the matter in both cases was
that although regulators were present, the systems had an infinite
number of fermions per unit Euclidean four-volume
\cite{neub_nar_plb1}. This led to
the chiral overlap. 

We spent several years testing and convincing ourselves that indeed
the unbelievable had happened and the lattice fermion problem had been
laid to rest\cite{neub_nar_npb1,neub_nar_npblong,twod_chiral}. Luckily, we
were not entirely alone doing this\cite{seif}.
Obviously, if the chiral gauge theory 
problem was solved one also had 
a way to exactly preserve global chiral symmetries in vector-like theories,
like QCD\cite{neub_nar_npblong}. How competitive this was 
when compared to standard numerical QCD was a question
we postponed. 
Eventually, focusing on QCD, a simplification of the overlap
in the case of vector-like gauge theories 
was found producing a relatively simple formula for the
fermion kernel for lattice Dirac fermions with exact global chiral
symmetry \cite{neub_plb_massless}.

Forgotten for many years was a prophetic paper by Ginsparg and Wilson in
which the nowadays well known GW relation 
for the kernel of a vector-like theory was written
down\cite{ginsp_wils}. Ginsparg and Wilson 
showed that if the kernel obeyed their
relation one would have both exact global chiral symmetries and
the $U(1)$ anomaly, directly on the lattice. They arrived
at their relation being motivated by
Renormalization Group ideas. In four dimensions a vector-like
gauge theory would be classically conformally invariant if matter
were massless, which in the fermion case also implies chiral symmetry.
Since the Renormalization Group is built to extract the anomalous realization
of dilations in the quantum context, it was a natural place to start.
The Renormalization Group provided them with a rather formal derivation
of their relation. To be sure, the GW relation itself, and its consequences were
however well defined and clear. But, they could not find
an explicit kernel satisfying their 
relation in the presence of a gauge background,
and probably because of this the idea was forgotten. 
It was a pleasant surprise to realize 
that the overlap Dirac operator relevant to the 
vector-like case satisfied the GW relation \cite{neub_prd,neub_plb_gw}. 
Of course, in
the context of the overlap construction it was obvious a priori
that one had global chiral symmetries and exactly massless quarks. 
But, that the GW relation would turn out to be satisfied was
not built in as an explicit requirement in the construction. 

Very soon after the overlap Dirac operator was shown to obey the GW
relation, Narayanan \cite{narayanan} showed that the derivation that
took one from the chiral overlap to the vector case by combining
a left handed fermion with a right handed one could be reversed.
Narayanan factorized the overlap Dirac operator using only
that the latter obeyed the GW relation and therefore it
became possible to start from the GW relation and get to the
chiral case by factorization. No new lattice 
results have been obtained starting from the GW relation directly,
although some derivations can be made to look more familiar. Later on, 
I shall explain in greater detail the connection between the overlap
and the GW relation. 

Let me note some important properties of the overlap approach to chiral
fermions. It works in any even dimension 
both for gauge and for gravitational
backgrounds. It is independent of the renormalizability
of the dynamics of the background. As
such, the overlap is not Renormalization Group motivated,
although Ginsparg and Wilson were. There is no meaningful GW relation
in odd dimensions. However, there are global issues in odd dimensional
gauge theories with massless fermions that are intimately related to
chiral fermions. These issues are captured in the overlap approach
which extends naturally to odd dimensions. 

\subsection{The basic idea}

\vskip .1cm

\begin{equation}
{\cal L}_\psi =\bar\psi \sla D \psi + \bar \psi (P_L {\cal M} +
P_R {\cal M}^\dagger )\psi
\end{equation}

$\bar\psi$ and $\psi$ are Dirac and the mass matrix ${\cal M}$ is
infinite. It has a single zero mode but its adjoint has no zero modes.
This were impossible if the mass matrix were finite. It clearly
means we have one massless Weyl fermions whose handedness can
be switched by interchanging the chiral projectors. It is very important
that as long as ${\cal M}{\cal M}^\dagger > 0$ this setup 
is stable under small deformations of the mass matrix. 
This stability comes from the internal
supersymmetric quantum mechanics generated by the mass matrix. 

Kaplan's domain wall suggests the following realization:
\begin{equation}
{\cal M} = -\partial_s - f(s)
\end{equation}
where $s\in (-\infty , \infty )$ and $f$ is fixed at $-\Lambda^\prime$
for negative $s$ and at $\Lambda$ for positive $s$. There is no
mathematical difficulty associated with the discontinuity at $s=0$.
Originally, the $s$-variable was discretized, but this is unnecessary. 

Before proceeding let me remark that other realizations of ${\cal M}$
ought to be possible, but nothing concrete has been worked out up to now.

The infinite path integral over the fermions is easily ``done'': on the positive
and negative segments of the real line respectively one has propagation with an
$s$-independent ``Hamiltonian''. The infinite extent means that at $s=0$ 
the path integrals produce 
the overlap (inner product) between the two ground states of the
many fermion systems corresponding to each side of the origin in $s$.
The infinite extent also means infinite exponents linearly proportional
to the respective energies - these factors are subtracted. One is
left with the overlap formula which expresses the chiral determinant
as $\langle v^\prime \{ U \} | v \{ U \} \rangle$. The states
are in second quantized formalism. By convention, they
are normalized, but their phases are left arbitrary. This
ambiguity is essential, as we shall see later on. It has no
effect in the vector-like case. 
In first quantized formalism the overlap is:
\begin{equation}
\langle v^\prime \{ U \} | v \{ U \} \rangle = \det _{k^\prime k} M_{k^\prime k}
\end{equation}
The elements of the matrix $M$ are 
the overlaps between single body wave-functions, $M_{k^\prime k}=
v^{\prime \dagger}_{k^\prime} v_k$. 
(In the 94 paper\cite{neub_nar_npblong} $M$ was denoted
by $O^{RR}$ and the single particle states $v_k , v^\prime_{k^\prime}$
by $\psi^{R+}_K , \psi^{R-}_{K^\prime}$.) The $v^\prime$'s span the negative
energy subspace of $H^\prime \sim \gamma_5 ( \sla D_4 +\Lambda^\prime )$
and the $v$'s span the negative
energy subspace of $H \sim \gamma_5 ( \sla D_4 -\Lambda)$. 
I used continuum like notation to emphasize that the Hamiltonians
are arbitrary regularizations of massive four dimensional Dirac
operators with large masses of opposite signs. One may wonder why
the different signs can at all matter. A simple way to see the difference
is to consider a gauge background consisting of one instanton.
While $\det H^\prime$ is positive, $\det H$ is negative. In a
complete, massive, four dimensional theory the mass sign could be traded
for a topological $\theta$ parameter and the two cases would correspond
to $\theta =0$ and $\theta=\pi$.  

The Hamiltonians only enter as defining the Dirac
seas and there is no distinction between the different levels within 
each sea; all that matters is whether a certain single particle state
has negative or positive energy. Thus, all the required information is
also contained in the operators $\epsilon =\varepsilon (H)$ and
$\epsilon^\prime = \varepsilon (H^\prime ) $ where $\varepsilon$ is the
sign function. Thus the $v^\prime$'s are all the $-1$ eigenstates of
$\epsilon^\prime$ and the $v$'s are all the $-1$ eigenstates of $\epsilon$.
To switch chiralities one only has to switch the sign of the Hamiltonians.
This is a result of charge conjugation combined with a particle-hole
transformation.

Nowadays the defining equations for  the states $v$ and $v^\prime$
are expressed with the
help of the projectors 
${{1+\epsilon}\over 2}$, ${{1+\epsilon^\prime}\over 2}$ (they annihilate
the states $v$ and $v^\prime$ respectively). But this is only notational
novelty.

When $\Lambda^\prime$ is taken to
infinity in lattice units one is left with $\epsilon^\prime =\gamma_5$.
Thus, $\epsilon^\prime$ becomes gauge field independent and so become the
associated states. (This simplification was already there in the Boyanowski,
Dagotto, Fradkin paper, but has been rediscovered in the domain wall context
by Shamir\cite{shamir}.) Physically, $\epsilon^\prime$ can be thought
of as a lattice representation of a continuum, positive
infinite mass, five dimensional Hamiltonian for 
Dirac fermions in a static gauge field;
one can decouple the fermions from the gauge field entirely. 
On the other hand, $\epsilon$ always maintains
a dependence on the gauge background and its trace
gives the gauge field topology. The parameter $\Lambda$
is restricted to a finite range
in lattice units and cannot be taken to infinity.
Physically, one can think
of $\epsilon$ as a lattice representation of a continuum, negative
infinite mass, five dimensional Hamiltonian for 
Dirac fermions in a static gauge field; the 
unavoidable dependence on the gauge field reflects the continuum result that
infinitely massive fermions in odd dimensions 
cannot decouple from the gauge fields
for both signs of the mass term.

\subsection {The vectorial case}

We add the states $w$ corresponding to the chirality opposite to that
represented by the states $v$ above. As just said, all this requires is
to switch the signs of the $\epsilon$'s. The left handed and right
handed fermions do not mix, each being self-coupled by $M^R_{k^\prime k}=
v^{\prime \dagger}_{k^\prime} v_k ,~~
M^L_{k^\prime k}= w^{\prime \dagger}_{k^\prime} w_k $. 

We wish to combine the two systems and get rid of the extra, unused dimensions
in each case. This is possible in the vector-like case, but not in
the chiral case, because, although the shapes of $M^R$ and $M^L$ change
as a function of the gauge background topology, they change in
a complementary way: The number of rows is fixed and the number of columns
of $M^R$ plus the number of columns of $M^L$ is also fixed, equal
to twice the number of rows. Thus $M^R$ and $M^L$ can be packed
together into a square matrix of fixed size. To describe $M^R$ or $M^L$
alone one needs a larger space because the shape of these matrices
fluctuates. To describe both matrices together however, extra dimensions
are not needed. 

In the most important case, at zero topology, 
we are searching for a simplified formula for the
product of the determinants of $M^R$ and $M^L$. The two $\epsilon$'s
generate a relatively simple algebra; the main new
operator in this algebra is the unitary operator $V=\epsilon^\prime \epsilon$. 
A very basic linear set of elements in that algebra
is $O=\alpha \epsilon +\beta \epsilon^\prime +\gamma 
\epsilon^\prime \epsilon +\delta$.
The matrix elements of $O$ between any $v$ or $w$ states are trivially
expressible in terms of corresponding overlaps. 
Picking
$\alpha=\beta=0,~\gamma=\delta={1\over 2}$ we can kill all $v-w$ cross
terms and the matrix elements of $O$ are determined by those of $M^R$
and $M^L$:
\begin{equation}
\pmatrix { w^\prime & v^\prime\cr }^\dagger {{1+\epsilon^\prime 
\epsilon}\over 2} \pmatrix {w & v\cr} = \pmatrix {M^L & 0\cr 0 & M^R \cr}
\end{equation}

Both $\pmatrix{w & v\cr}$ matrices are unitary since
the columns make up orthonormal bases. Using charge conjugation one can assure
that the determinants of the two $M$-matrices are complex conjugate of each
other. With
\begin{equation}
D_o ={{1+\epsilon^\prime \epsilon }\over 2}
\end{equation}
one trivially derives $\det D_o = |\det M^L |^2$. 
When $\Lambda^\prime =\infty$, $\pmatrix{w^\prime & v^\prime\cr}$ is the unit
matrix. Moving the unitary factor $\pmatrix{w& v\cr}$ to the other side
of equation (4) we see that $D_o$ has been ``factorized''. The columns
$v$ span the kernel of ${{1+\epsilon}\over 2}$ and the 
columns $w$'s span the orthogonal
complement of this subspace. 

Another way to decouple $v$ from $w$ is to choose in $O$
$\alpha=\beta={1\over 2},~\gamma=\delta=0$:
\begin{equation}
\pmatrix { w^\prime & v^\prime\cr }^\dagger {{\epsilon^\prime +  
\epsilon}\over 2} \pmatrix {w & v\cr} = \pmatrix {M^L & 0\cr 0 & - M^R \cr}
\end{equation}
At $\Lambda^\prime =\infty$, and after moving the 
unitary factor $\pmatrix{w& v\cr}$ to the other side
of equation (6) we obtain the hermitian overlap Dirac 
operator, $H_o =\epsilon^\prime D_o$ studied at SCRI\cite{scri}. 

It is important that even if we keep $\Lambda^\prime$
finite in lattice units and $H^\prime$ has a nontrivial gauge dependence, 
$H^\prime$ always has exactly as many
negative as positive energy eigenstates and there is an impenetrable
(as a function of the gauge background) gap in its spectrum around zero.
In other words, $\epsilon^\prime$ is never sensitive to
gauge field topology. 

Let me add here that a version of $D_o$ can be derived starting with a lattice
implementation of the see-saw mechanism obeying a Froggatt-Nielsen
symmetry. This $D_o$ is obtained in the limit of infinite see-saw partners
\cite{neub_prd}.

\subsection{Topology and fermions}

The topological charge $Q\{ U\}$ is the difference between the
number of columns and rows of $M^L$, which is the negative of
the same quantity for $M^R$. Since $tr (\epsilon^\prime ) \equiv 0$,
$Q\{ U \} = {1\over 2} tr \epsilon$. When $Q\{ U \} \ne 0$, $\det
\pmatrix {M^L &0\cr 0& M^R \cr} \equiv 0$ because either among the first
columns or among the last there are too many zeros to maintain
linear independence. This implies $\det D_o =0$, and hence
exact zero modes for the overlap Dirac operator \cite{neub_nar_prl}. 
Using $tr\epsilon^\prime =0$ the formula $Q\{ U \} = {1\over 2} tr \epsilon$
can be written in many equivalent 
ways. These days a popular way is  
$Q\{ U \} = tr \epsilon^\prime D_o$ with the 
sum over sites contained in the trace made explicit. The summand is a lattice
version of the topological density.

The impact of topology on fermion dynamics is
easiest to see in second quantized language (our
original formulation). We denote second quantized operators
by hats: $\hat H = \hat a^\dagger H \hat a,~~\hat H^\prime = 
\hat a^\dagger H^\prime \hat a ,~~\hat N = \hat a^\dagger \hat a$,
$[\hat H, \hat N ]= [\hat H^\prime ,\hat N ]=0$. The second 
quantized states entering the overlap satisfy: $\hat N |v^\prime \rangle
={1\over 2} {\cal N} |v^\prime \rangle$ and 
$\hat N |v^\prime \rangle =({1\over 2} {\cal N} +Q\{ U\} ) 
|v^\prime \rangle$. Here ${\cal N}=tr 1$. Clearly, $Q\{ U \} \ne 0$ 
forces $\langle v^\prime | v \rangle =0$. This immediately 
leads to nonvanishing, automatically normalized 't Hooft vertices. 
For example, if $Q\{ U \} =1$, $\langle v^\prime| \hat a | v\rangle \ne 0$.

The consequences of the 't Hooft vertices are far reaching. 
In the vector-like case they provide the solution to the $U(1)_A$ 
problem, now in an entirely rigorous setting. In a background 
that carries topological charge 1 for each flavor we shall have 
$\langle v^\prime |\hat a | v \rangle \ne 0,~~ 
\langle w^\prime | \hat a^\dagger | w \rangle \ne 0$ which gives the 
two $\bar\psi_R \psi_L$, $\bar\psi_L \psi_R$ factors per 
flavor that make up the vector-like 't Hooft vertex.  
In the chiral case one can get explicit fermion number violation. 
A simple example of this can be found in two dimensions, in 
an abelian gauge model with fermionic matter consisting of one 
charge 2 right handed fermion $\psi$ and four left handed charge 
-1 fermions $\chi_\alpha~~(\sigma_\mu =(1,i))$. 
The 't Hooft vertex gives a nonzero 
expectation value to the operator $V=\psi \sigma\cdot\partial
\psi \chi_1\chi_2\chi_3\chi_4$. The model is exactly soluble and 
known to have a massless composite sextet of fermions $\Phi_{\alpha\beta} = 
\psi\chi_\alpha\chi_\beta$. These fermions are actually Majorana-Weyl, 
although the original fermions were just Weyl. This is needed 
to match the global $SU(4)$ anomalies associated with the four 
$\chi$ fermion fields at the composite level. The 't Hooft vertex $V$ 
(note the absence of barred fermion fields) provides the kinetic 
energy term for these composites. 

The confirmation of the above features numerically 
in the 21111 chiral model represented a major step since it showed that 
even the subtler details of chiral fermion dynamics 
were captured by the overlap in an effortless way. More
traditional approaches to the chiral fermion problem always
had to come up with tenuous explanations for how
such effects might be recovered in the continuum. 

\subsection{Chiral symmetry breaking}

We find ourselves on the lattice, with exact global chiral 
symmetries, with correct anomalies and with exactly obeyed 
mass inequalities. Based on the continuum it seems that the 
day is not far where we shall be able to claim to have a 
rigorous proof of spontaneous chiral symmetry breaking 
directly on the lattice. 

In the meantime, numerical work on the spectral properties of $H_o$ 
by SCRI has produced ``experimental'' evidence for spontaneous chiral symmetry breaking. 
Related observations were made by the Kentucky group\cite{scri,kfl}. 
See also \cite{lelo}.

\subsection{Anomalies}

The second quantized states entering the overlap each come from 
a single body Hamiltonian which is analytic in the gauge link variables. 
Thus, they carry Berry phases. There is a natural connection 
(an abelian gauge field) over the space of gauge fields (playing 
the role of parameters in the familiar Berry setup). Integrating Berry's 
connection along a smooth closed loop in gauge field space generates 
an invariant phase, Berry's phase. There is such a connection 
associated with each state in the overlap, but the one associated 
with $|v^\prime \rangle$ can be made to vanish by taking $\Lambda^\prime$ 
to infinity. ${\cal A} = \langle v\{ U \} | d v \{ U \} \rangle $ is 
the expression giving the connection once some representatives of the 
rays $| v\{ U \} \rangle $ are chosen (possibly 
using patches with overlays). 
${\cal A}$ is not quite a function over the space of gauge fields; 
it is a connection, in the sense that it could be defined in patches 
and in their overlays the several definitions could 
differ by gauge transformations. 
Berry's phase is nontrivial at the ``perturbative level'' because 
${\cal A}$ has curvature (abelian field strength) ${\cal F}= d{\cal A}$ 
which does not vanish. Unlike ${\cal A}$, ${\cal F}$ is a function 
of the gauge field independent of the ray representatives used 
to define the connection. In second quantized notation ${\cal F} = 
\langle d v \{ U \} | d v\{ U \} \rangle$ (antisymmetrization is 
implicit). In first quantized 
language one has ${\cal F} = \sum_{k\in{\rm Dirac~ sea}} dv_k^\dagger dv_k$. 
Note that the overlaps entering the connection and the curvature 
only compare variations to the same state. This is why ${\cal F}$ 
is a local functional of the gauge background. It does not depend 
on phase choices, so can be expressed in terms of the sign function 
alone: ${\cal F} = -{1\over 4} tr\left ( 
\varepsilon(H)d\varepsilon(H)d\varepsilon (H)\right )$. 

A special case is interesting. Take the gauge group as $SU(2)$
and the fermions in the fundamental representation. There is
a simple basis in which $H\{ u \}$ is a real matrix with no
complex entries. Thus, it is natural to choose all eigenstates 
real and therefore Berry's connection and its curvature vanish. 
Nevertheless, Berry's phase factors can be nontrivial giving 
sign flips when states are taken round some loops. The $U(1)$ 
bundle of states one usually has in the overlap is replaced by 
a $Z_2$ bundle, and the latter could be twisted. This is how 
Witten's global anomalies show up in the overlap
\cite{neub_z2}. Let me remind 
you that seeing Witten's global anomalies was beyond the reach 
of all approaches to the chiral fermion problem before the overlap.

My main message if that Berry's phase encodes all anomalies in the 
theory. Let us see how this works in the ordinary, complex case
\cite{neub_geom}.

In the continuum one defines two currents. (I shall restrict my 
discussion to a nonabelian semisimple gauge group and to a spacetime 
topology of a $d$-torus.) The consistent current is the variation 
of the chiral determinant with respect to the gauge field. As a variation 
it obeys ``curl-free'' constraints, also known as the Wess-Zumino 
conditions. When there are anomalies the consistent current 
cannot be covariant with respect to gauge transformations. If the 
determinant were gauge invariant the consistent current would 
trivially also be covariant. Even when there are anomalies and 
the consistent current cannot be covariant it can be 
made so by adding a local, exactly known polynomial in the 
gauge fields and their derivatives. This quantity is called $\Delta J$.
Although both the covariant and the consistent currents 
are non-local functionals (in the continuum) of the gauge background, 
their difference, $\Delta J$, is local. $\Delta J$, by itself, 
fixes the anomaly. In short, gauge invariance is restorable 
if and only if $\Delta J$ vanishes (on account of anomaly cancelation).
 
Now let us go back to the lattice. We make some smooth phase choice for 
the states representing the ground state rays and compute 
the variation of the overlap with respect to 
the external gauge fields. This should produce a lattice version of the  
consistent current on the lattice because it is the variation 
of something. One writes $J^{\rm cons} = {\cal A}-{\cal A}^\prime
+ J^{\rm cov}$. The Berry phase terms contain the part 
of $J^{\rm cons}$ which is not guaranteed to be gauge
covariant. The remainder, defined as $J^{\rm cov}$ is given 
(at $\Lambda^\prime =\infty$)
by ${{\langle v^\prime | d v \rangle_{\perp}}\over
{ \langle v^\prime |  v \rangle}}$; it 
is independent of the phase choices and has naive 
gauge transformation properties. Here, $| x \rangle_\perp \equiv 
| x \rangle - \langle v | x\rangle | v\rangle $, 
which is independent of the phase of $|v \rangle $.  
The ``curl'' of $\Delta J = {\cal A}-{\cal A}^\prime $
is ${\cal F} - {\cal F}^\prime$ and does not vanish in general.

The analogy sketched above has not been yet fully fleshed out,
but one result is available. 
Pick an abelian background in the direction of  
a $U(1)$ subgroup with charges $q_i$. 
In the abelian context, ${\cal F}$ is gauge invariant and
can be viewed as defined over gauge orbits. 
If the anomaly does not vanish one
can find a two-torus in the space of gauge orbits over which
the necessarily quantized integral of ${\cal F}$ is nonzero.
(In $d$ even dimensions the integral $\int {\cal F}$ 
goes as $\sum_i q_i^{{d\over 2}+1}$). This implies
that no ``small'' deformation of $H\{ U \}$ can make ${\cal F} \equiv 0$
and hence $\Delta J \ne 0$. This leads to 
two conjectures (for the complex
case):

If and only if anomalies cancel it is possible to smoothly deform
$H\{ U \}$ and $H^\prime \{ U \}$ such that ${\cal F}={\cal F}^\prime$.
If  ${\cal F}={\cal F}^\prime$ one can choose the second quantized states
$| v\{ U \} \rangle$ and $| v^\prime \{ U \} >$ smoothly such that
the action of the gauge group is non-projective: for any $g\in {\cal G}$
$G(g) | v \{ U \} \rangle > = | v\{ U^g \} \rangle $ and the same for
$| v^\prime \rangle$. If these conjectures prove true one can preserve
exact gauge invariance of the overlap  if anomalies cancel, but, 
to do so, one needs to fine tune the Hamiltonians. 

So, we must ask whether this is ``natural''. The answer is that fine
tuning is not necessary to get full gauge invariance in the continuum
limit. Even before fine tuning the gauge breaking of the overlap
is of a specific kind because the Hamiltonians are gauge covariant,
implying (excluding backgrounds with degenerate fermionic ground states)
$G(g) | v \{ U \} \rangle = e^{i\chi (\{ U \}, g ) } | v \{ U^g \} \rangle $.
$\chi - \chi^\prime$ is a lattice Wess-Zumino action. By fine tuning we
conjectured that one can make $\chi= \chi^\prime$ if anomalies
cancel. But even if the lattice
Wess-Zumino action is not zero, as long as anomalies 
cancel, it can be small in the sense that one can expand in it. 
(The cancelation of anomalies implies that in the continuum limit
the lattice Wess-Zumino action will have no contribution from the
continuum Wess-Zumino action.) Then
the mechanism discovered by F{\" o}rster, Nielsen and Ninomiya shows that
exact gauge invariance will be restored in the continuum limit and the
Higgs like degrees of freedom representing gauge transformations decouple
\cite{FNN}. 
This was checked numerically in the above mentioned abelian two dimensional
model already in 1997. Although this was only two dimensions it was not at all 
trivial. 

The next chiral speaker will concentrate on 
the phase of the overlap \cite{luscher}. Berry's connection and
curvature will be seen to play a central role. 
It is important to stress that the problem
has reduced to a phase choice only because in the overlap 
this is the single source
of gauge breaking, just as emphasized in the continuum
context by Fujikawa: Any fermionic correlation
function in a fixed gauge background violates gauge covariance 
by no more and no less than the determinantal anomaly. 

\subsection{ GW and overlap}

We already heard that the chiral overlap produced the vector-like 
operator $D_o$ and that $D_o$ can be factorized to give
back the chiral overlap. Now we focus on the relation between $D_o$ and
GW. For related considerations, see \cite{twchiu}.

Let us first specify what is meant by GW (I adopt
a restricted definition including $\gamma_5$-hermiticity.):  
(1a) One is given a local hermitian positive operator $R$
which commutes with $\gamma_5$. The issue is to find a Dirac operator 
satisfying $\{ \gamma_5, D^{-1} - R\} =0$ and (1b) $\gamma_5 D = (\gamma_5 D)^\dagger$.

Clearly, the operator $D_c^{-1} = D^{-1} - R$ anticommutes with 
$\gamma_5$ and is $\gamma_5$-hermitian. (By standard wisdom, $D_c$ cannot
be local.) Define the operator $V = {{1+D_c}\over {1-D_c}}$. $V$ is
seen to be $\gamma_5$-hermitian and unitary. Inverting the relation,
we find $D_c^{-1} = {{1-V}\over {1+V}}$ leading trivially to
$D=(1+V) {1\over {1+R - (1-R) V}}$ which is the most general solution
of (1a+b), in terms of a unitary hermitian 
operator $\epsilon \equiv \gamma_5 V$. ($D_o$ corresponds to $R=1$.) 
Obviously, $\epsilon$ squares to unity. Although
this satisfies the GW requirement, it is not enough to produce
massless fermions. One also needs
that topology be given by $Q\{ U \} = {1\over 2} tr \epsilon$. In our
realization of the overlap we used $\epsilon = \varepsilon (H)$ with
a sparse $H\{ U \}$ analytic in the link variables and showed that
topology and perturbation theory produce the correct chiral answers.

It is trivial that the overlap provides a solution to GW.
What is the physical meaning of $\epsilon$ ? 

Physically, $\epsilon$ by itself describes Dirac fermions, but they have
infinite mass. Therefore, unlike $D_c$, $\epsilon$ is local (except
when ill defined). In the continuum, the infinite mass Hermitian Euclidean
Dirac operator would have a spectrum concentrated at $\pm \infty$. 
$\epsilon$ is a rescaled version, with spectrum at $\pm 1$. Any lattice
operator $H$, representing Dirac fermions with order ${1\over a}$ negative
mass produces an $\epsilon = \varepsilon (H)$. A solution of GW, $\epsilon$,
that also satisfies the additional conditions required of massless fermions
is an acceptable $H$, and reproduces itself in an overlap construction since
$\epsilon =\varepsilon (\epsilon )$. The overlap provides more flexibility,
allowing the replacement of $\gamma_5$ by $\epsilon^\prime$.

It is
unreasonable to view the GW relation as pivotal in Nature because it is
just a formula, not the embodiment of a fundamental principle. Moreover,
the formula accepts also unphysical solutions. 
On the other hand, the overlap
is a direct reflection of a system consisting of an infinite number
of fermions governed by some internal dynamics realizing an internal index; 
it is easier to accept
that this is a natural mechanism, conceivably operative in Nature. 

Had events in this decade occurred in reversed chronological 
order the infinite
fermion number ``explanation'' of the GW relation might have been
viewed as an inspired insight. 

\section{A list of projects}

There is plenty to do and you are invited
to join the chiral subfield! 
To make my case, I shall present a list of projects. I
don't suggest that you slavishly 
execute any one of them. The intention is more to 
inspire you, so you come up with your own
idea. The main project seems difficult to me: 

${\bf \star}$ Find a genuinely non-overlap way to solve the chiral
fermion problem. If this is possible we shall conclude 
that the overlap only solved
our problem, not necessarily that faced by Nature. 

Let me turn to less ambitious proposals:

\subsection{General particle physics}

$\bullet$ Examine which aspects of low energy physics would be particularly
sensitive to an UV regulator of the overlap type.

$\bullet$ Find a natural way to explain the subtraction 
of the infinite Dirac sea
vacuum energies.

$\bullet$ Prove that one cannot find an acceptable 
solution to the GW relation which is nearest neighbor even 
in only one direction. Proceed to argue 
that unitarity in Minkowski space requires an
infinite number of fermions.

\subsection{Numerical 4D chiral gauge theories}

In order to avoid dealing with a complex measure but still
treat a non-trivial chiral model I suggest to solve numerically 
an $SU(2)$ gauge theory with one Weyl $j={3\over 2}$
multiplet. The chiral determinant is real and there are no Witten anomalies.
But, there also is no singlet $\psi -\psi$ bilinear.

$\bullet$ What is the phase structure as a function of $\beta$, 
the gauge coupling ?

$\bullet$ Does the model confine ? What is its particle spectrum ?

$\bullet$ Are there massless fermion states ?

\subsection{QCD}

$\bullet$ Go to the $F_4$ lattice to disallow terms 
of the form $\sum_\mu p_\mu^4 $
which are scalars on a hypercubic lattice but not in nature. (This is analogous
to Higgs work, where, strictly speaking, claims about Nature on the basis of
lattice work cannot be made using hypercubic lattices without fine tuning away
the $\sum_\mu p_\mu^4 $ term.) Compute, by Monte Carlo simulation, 
the order $p^4$ coefficients
in a chiral effective Lagrangian for pions resulting from 
massless quarks  \cite{rebbi}. 
It is suggested to do this using finite size soft
pion theorems of the type used previously in $F_4$ lattice 
Higgs work \cite{soft}. (Let me take the opportunity to correct a misunderstanding
that occurred during the discussion following my talk; contrary to a comment
from the audience, this problem has not been solved in a poster presented
at this conference \cite{lelo}.)

$\bullet$ Use $D_o$ to define nonperturbative improvement coefficients to
standard actions. The improvement intends to hasten the restoration
of chirality in the
continuum limit. On a gauge configuration typical of a fixed
$\beta$ evaluate $c$ and $c^\prime$ by minimizing $|| c^\prime D_o - (D_W + c
\sigma\cdot F ) ||^2$.

$\bullet$ The sign function $\varepsilon (M)$ is well defined
even for complex $M$: investigate QCD at nonzero chemical potential but
zero quark mass using the overlap.

$\bullet$ Use the Wilson-Dirac operator $D_W$ to define the pure gauge
action, as well as $\epsilon$. This should reduce the density of
states with low $H_W^2$ eigenvalues. For example, a pure gauge action
could contain the term $tr{1\over{H_W^2}}$, or,  
alternatively, one could use the determinant of a function 
of $H_W$ implemented by auxiliary heavy bosonic fields.

\end{document}